\documentclass[journal, draftcls, one column, 12pt]{IEEEtran}
\usepackage{amsmath}
\usepackage{graphicx}
\usepackage{amssymb}
\usepackage{cases}
\usepackage{color}
\usepackage{cite}
\usepackage[ruled,vlined,lined,ruled,linesnumbered]{algorithm2e}
\usepackage{relsize}
\usepackage[nodisplayskipstretch]{setspace}
\usepackage[top=1in, bottom=1in, left=0.7in, right=0.7in]{geometry}

\begin{document}
\title{Wireless Powered Communication: Opportunities and Challenges}
\author{Suzhi~Bi, Chin Keong Ho, and Rui Zhang}
%
\maketitle

\vspace{-1.8cm}

\section*{Abstract}
The performance of wireless communication is fundamentally constrained by the limited battery life of wireless devices, whose operations are frequently disrupted due to the need of manual battery replacement/recharging. The recent advance in radio frequency (RF) enabled wireless energy transfer (WET) technology provides an attractive solution named wireless powered communication (WPC), where the wireless devices are powered by dedicated wireless power transmitters to provide continuous and stable microwave energy over the air. As a key enabling technology for truly perpetual communications, WPC opens up the potential to build a network with larger throughput, higher robustness, and increased flexibility compared to its battery-powered counterpart. However, the combination of wireless energy and information transmissions also raises many new research problems and implementation issues to be addressed. In this article, we provide an overview of state-of-the-art RF-enabled WET technologies and their applications to wireless communications, with highlights on the key design challenges, solutions, and opportunities ahead.

\section{Introduction}
Limited device battery life has always been a key consideration in the design of modern mobile wireless technologies. Frequent battery replacement/recharging is often costly due to the large number of wireless devices in use, and even infeasible in many critical applications, e.g., sensors embedded in structures and implanted medical devices. RF-enabled wireless energy transfer (WET) technology provides an attractive solution by powering wireless devices with continuous and stable energy over the air. By leveraging the far-field radiative properties of electromagnetic (EM) wave, wireless receivers could harvest energy remotely from the RF signals radiated by the energy transmitter. RF-enabled WET enjoys many practical advantages, such as long operating range, low production cost, small receiver form factor, and efficient energy multicasting thanks to the broadcast nature of EM wave.

\begin{figure}
\centering
  \begin{center}
    \includegraphics[width=0.9\textwidth]{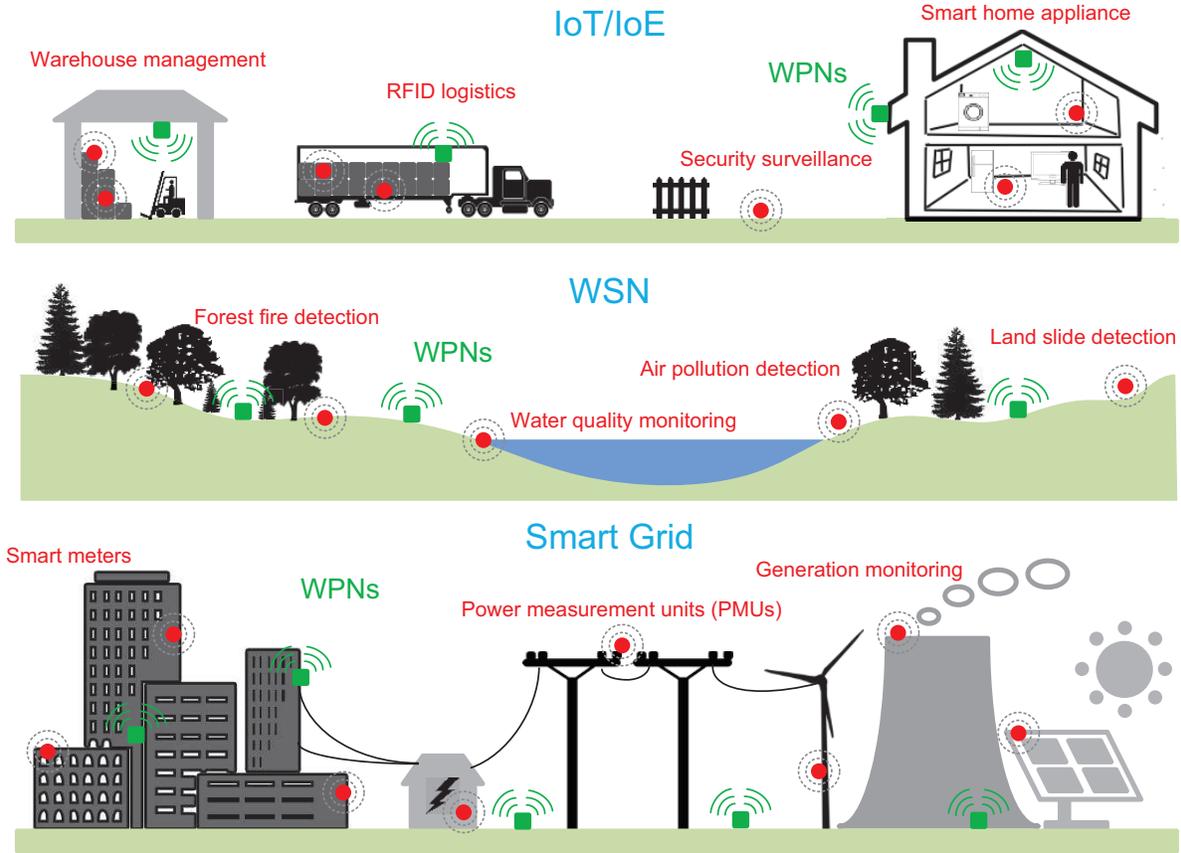}
  \end{center}
  \caption{Example applications of WPC in IoT/IoE systems, WSNs for environment monitoring, and smart power grid. The green nodes denote the wireless power nodes (WPNs), which transmit RF energy to the wireless powered devices denoted by red nodes in the figure.}
  \label{71}
\end{figure}

One important application of RF-enabled WET is \emph{wireless powered communication} (WPC), where the wireless devices use the harvested RF energy to transmit/decode information to/from other devices. Without being interrupted from energy depletion due to communication usage, WPC is expected to improve user experience and convenience, with a higher and more sustainable throughput performance than conventional battery-powered communication. WPC can also be applied in sensors with much lower maintenance cost and enhanced flexibility in practical deployment. Due to the high attenuation of microwave energy over distance, RF-enabled WET is commonly used for supporting low-power devices, such as RFID (radio-frequency identification) tags and sensors. However, the recent advances of antenna technologies and RF energy harvesting circuits have enabled much higher microwave power to be efficiently transferred and harvested by wireless devices \cite{2002:Suh}. Therefore, we envision in Fig.~$\ref{71}$ that WPC will be an important building block of many popular commercial and industrial systems in the future including the upcoming internet of things/everything (IoT/IoE) systems consisting of billions of sensing/RFID devices as well as large-scale wireless sensor networks (WSNs). We also envision that RF-enabled WET is key component of the ``last-mile" power delivery system, with the smart electrical power gird forming the ``backbone" or core power network.

Before proceeding to the discussion of RF-enabled WET/WPC, it is worth pointing out its relation to another green communication technique, namely, \emph{energy harvesting} (EH), where wireless devices harness energy from energy sources in environment not dedicated to power the wireless devices, such as solar power, wind power, and ambient EM radiations. Unlike RF-based EH from ambient transmitters, the energy source of WET is stable, and more importantly, fully controllable in its transmit power, waveforms, and occupied time/frequency dimensions, to power the energy receivers. With a controllable energy source, a \emph{wireless powered communication network} (WPCN) could be efficiently built to power multiple communication devices of different physical conditions and service requirements. Besides, with RF-enabled WET, information could also be jointly transmitted with energy using the same waveform. Such a design paradigm is referred to as \emph{simultaneous wireless information and power transfer} (SWIPT), which is proved to be more efficient in spectrum usage than transmitting information and energy in orthogonal time or frequency channels \cite{2013:Zhang,2013:Zhou}.

In this article, we first provide a brief overview of state-of-the-art RF-enabled WET technologies. Then, we focus on introducing RF-enabled WPC in the following three topics:
\begin{itemize}
  \item the circuit model and advanced signal processing techniques used for WET;
  \item the design tradeoffs in joint energy and information transmission for SWIPT;
  \item the design challenges and opportunities in WPCN.
\end{itemize}
At last, we discuss the future research directions of WPC and conclude the article.

\section{State-of-the-art of RF-enabled WET}
Although WET has gained popularity in the recent years, it is in fact a technology under development for more than a century (see \cite{2011:Shinohara} for the detailed historical development of WET). The existing WET technologies could be categorized into three classes based on the key physical mechanisms employed: inductive coupling, magnetic resonant coupling and EM radiation. Among them, the first two types exploit the non-radiative near-field EM properties associated with an antenna for short-range high power transfer. Currently, inductive coupling WET is well-standardized, with applications such as charging mobile phone and medical implanted devices. However, due to the drastic drop of magnetic induction effect over distance, inductive coupling typically operates within only several centimeters range. The operating range of  magnetic resonant coupling WET could be as large as few meters. However, to maintain the resonant coupling, the receiver could not be flexibly positioned as it is optimized for some fixed distance and circuit alignment settings. Besides, transmitting energy to multiple receivers is challenging as it requires careful tuning to avoid interference due to mutual coupling effect.

On the other hand, RF-enabled WET exploits the far-field radiative properties of EM wave to power wireless devices over moderate to long distance. A typical RF-powered RFID tag could be powered at $4$ meters away (with around $0.5$ mW received RF power), and some RF energy harvesting chips have a maximum $12$-$14$ meters line-of-sight operating radius (with around $0.05$ mW received RF power).\footnote{Please refer to the website of Powercast Corp. (http://www.powercastco.com) for detailed product specifications.} In practice, RF-enabled WET uses inexpensive RF energy receivers, which could be flexibly positioned and made very tiny to fit into commercial devices. Besides, transmitting energy to multiple receivers is easily achieved with the broadcasting property of microwaves. The major constraint of the application of RF-enabled WET is the high attenuation of microwave energy over distance. Nonetheless, with the continuing decrease of device operating power (as low as a few microwatts for some RFID tags), and the recent application of MIMO technology that significantly enhances the wireless energy transfer efficiency, we could expect more and more important applications of RF-enabled WET in the future.

\section{Network Model for Wireless Powered Communication}

\begin{figure}
\centering
  \begin{center}
    \includegraphics[width=0.7\textwidth]{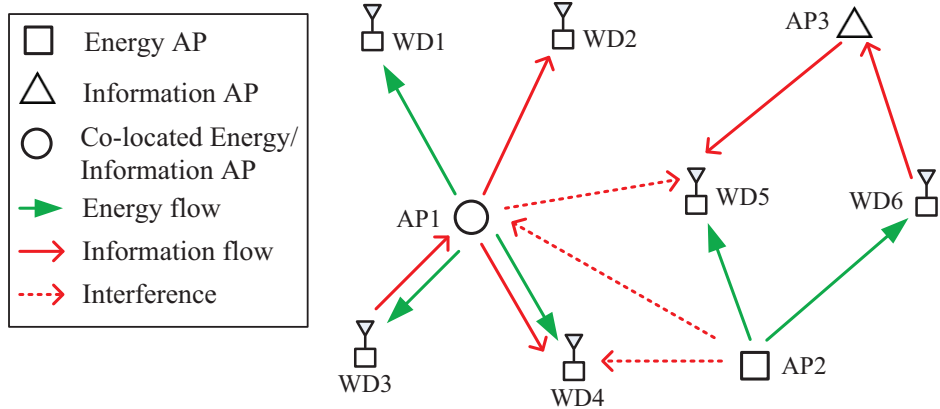}
  \end{center}
  \caption{A network model for wireless powered communication.}
  \label{72}
\end{figure}

In Fig.~$\ref{72}$, we present a network model to illustrate the basic concepts of WPC. In the downlink (DL), WET-enabled energy access points (APs) with stable power supply (e.g., AP2 in Fig.~$\ref{72}$) transmit energy to a set of distributed wireless devices (WDs). Meanwhile, the WDs could use the harvested energy to transmit/receive information to/from the information APs (e.g., AP3 in Fig.~$\ref{72}$) in the uplink (UL) and downlink, respectively. Besides, energy and information APs could be integrated into a co-located energy/information AP (e.g., AP1 in Fig.~$\ref{72}$), which both transmits energy and provides data access to the WDs. In particular, three canonical operating modes are specified as follows:
\begin{itemize}
  \item WET: energy transfer in the DL only, e.g., AP1 to WD1 and AP2 to WD5;
  \item SWIPT: energy and information transfer in the DL, e.g., AP1 to WD4;
  \item WPCN: energy transfer in the DL and information transfer in the UL, e.g., AP1 to WD3.
\end{itemize}
Accordingly, the WDs all perform energy harvesting (EH) in the WET mode, with applications such as charging sensors for sensing operations. Additionally, the WDs perform EH in the DL transmission of WPCN mode while sending data in the UL with harvested energy, with applications such as sensor battery charging and data collection in a WSN \cite{2014:Ju1}. For the SWIPT mode, the WDs perform both EH and information decoding (ID) in the DL with the same received signals, each using harvested energy to power its information decoder, e.g., in an energy-self-sustainable information broadcast network \cite{2013:Zhang,2013:Zhou}. In practice, a WPC network could also include other general network models consisting of multiple co-located or separated information/energy transmitters, and receivers with heterogeneous operating modes. For instance, WD6 harvests energy from an energy AP (AP2) and transmits data to a different information AP (AP3); AP1 transmits energy and information respectively to two separated energy and information receivers at the same time, i.e., WD1 and WD2. Besides, energy transmission could potentially generate interference to the information receivers operating in the same frequency band (see the red dashed lines in Fig. $\ref{72}$), for which a joint design of energy/information transmissions is highly desired. In the following sections, we will focus our discussions on the three canonical operating modes, i.e., WET, SWIPT and WPCN, respectively.

\section{Wireless Energy Transfer}
In this section, we introduce the RF energy receiver structure and advanced signal processing methods to enhance the energy transfer efficiency of RF-enabled WET.
\subsection{RF energy receiver model}
\begin{figure}
\centering
  \begin{center}
    \includegraphics[width=0.6\textwidth]{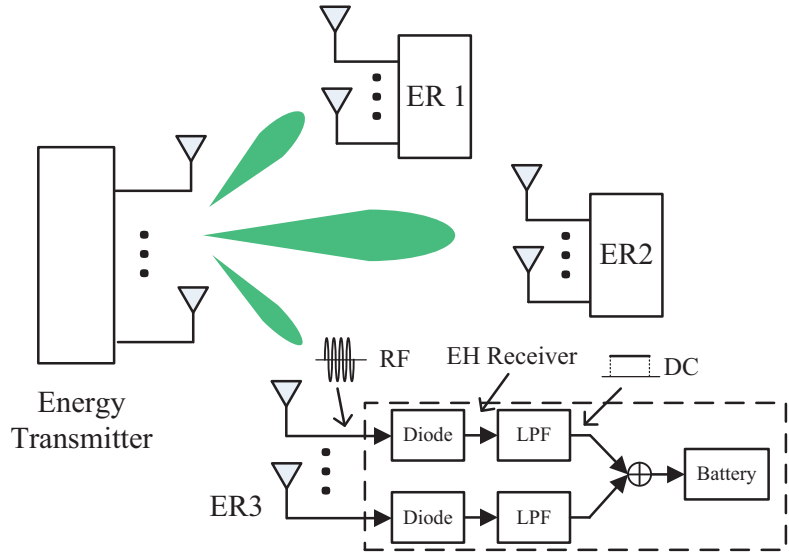}
  \end{center}
  \caption{A wireless energy broadcast network and the energy receiver structure. At ER $3$, the received RF signal is converted to a DC to charge a battery.}
  \label{73}
\end{figure}

We consider in Fig.~$\ref{73}$ an energy transmitter transferring RF energy to multiple energy receivers (ERs), where each transmitter/receiver is equipped with multiple antennas in general. The transmitted energy signals are in general modulated signals, e.g., pseudo-random signals, instead of an unmodulated sinusoid tone that is used widely in practice. The signals could be designed to avoid a spike in the power spectral density (PSD), and to satisfy the PSD requirement for safety and interference considerations. Therefore, wireless energy transmission occupies a certain bandwidth similar to information transmission, determined by the modulated baseband signal.

The RF energy harvesting (EH) circuit model is also depicted for ER $3$ in Fig.~$\ref{73}$. The EH receiver is based on a \emph{rectifying circuit}, consisting of a diode and a passive low-pass filter (LPF), which converts the received RF signal to a DC (direct current) signal to charge the built-in battery to store the energy. By the law of energy conservation, the harvested energy per unit symbol time at an ER, denoted by $Q$, is proportional to the received RF power $P_r$, i.e.,
\begin{equation}
Q = \eta \cdot P_r = \eta \cdot P_t \cdot D^{-\alpha} \cdot G_A.
\end{equation}
Here, $0 < \eta < 1$ denotes the overall receiver energy conversion efficiency, $P_t$ denotes the transmit power, $D$ denotes the distance (normalized with respect to a given reference distance) from the ER to the transmitter, $\alpha\geq 2$ denotes the path loss factor and $G_A$ denotes the combined antenna gain of the transmit and receive antennas. For instance, using two antennas at both the energy transmitter and the receiver, we could achieve a beamforming gain to increase the harvested energy by about $4$ times ($6$ dB) compared to the case with single-antenna transmitter and receiver with the same transmitter power. This could be more cost-effective in practice than the alternative approach of improving the energy conversion efficiency $\eta$ (say, from $25\%$ to $99\%$) at the receiver with more sophistically designed rectifying circuits.

\subsection{Energy beamforming}
Using multiple antennas not only provides antenna power gain as stated above, but also enables advanced \emph{energy beamforming} (EB) techniques to focus the transmit power in a smaller regions of space to bring significant improvement to the energy transfer efficiency \cite{2013:Zhang}. By carefully shaping the transmit waveform at each antenna, EB could control the collective behavior of the radiated waveforms, such that they are combined coherently at a specific receiver, but destructively at the others. In general, the larger number of antennas installed at the energy transmitter, the sharper the energy beam that could be generated towards a particular spatial direction. With only one ER, the transmitter could steer a single sharp beam to maximize the harvested energy. When there are multiple ER as in Fig.~$\ref{73}$, however, generating a single beam may result in severe unfairness among the receivers, also known as the (energy) near-far problem, where users near the transmitter harvest much more energy than the far users. In this case, the transmitter may need to generate multiple energy beams towards different directions to balance the energy harvesting performance among the receivers \cite{2014:Liu}.

An efficient EB design requires availability of accurate knowledge of the channel state information at the transmitter (CSIT). However, this is often difficult to achieve in practice. On one hand, many simple energy receivers have no baseband signal processing capability to perform channel estimation. On the other hand, accurate channel estimation consumes significant amount of time and energy, which may offset the energy gain obtained from a refined EB. Besides, receiver mobility could cause time-varying channels, making channel tracking difficult.

Various efficient channel estimation methods have been proposed to perform EB under imperfect CSIT by exploiting the received energy levels over time and balancing the tradeoff between energy consumption and EB gain (see, e.g., \cite{2014:Zeng} and the references therein). Besides, robust EB design could be pursued to generate energy beams based on the statistical knowledge of CSIT. Another promising method is to perform EB using distributed antennas. This effectively reduces the amount of feedback signals for channel estimation, as a receiver harvests energy only from a small subset of nearby transmitting antennas. Besides, the deployment of distributed antennas also reduces the range between the energy transmitters and receivers, thus is effective to solve the near-far problem caused by using a single energy transmitter. In this case, however, efficient coordination among the distributed antennas is needed.

\section{Simultaneous Wireless Information and Power Transfer}
When WET is applied to power communication devices, it will inevitably occupy part of the spectrum used for communication purpose. To avoid the interference to communication, a simple but spectrally inefficient method is to transmit energy and information in orthogonal frequency channels. Alternatively, SWIPT designs seek to save the spectrum by transmitting information and energy jointly using the same waveform. This is intuitively achievable, as any waveform for information transmission also carries energy to be harvested by the same or different receivers. However, an efficient SWIPT scheme involves a rate-energy tradeoff in both the transmitter and receiver designs to balance the information decoding (ID) and energy harvesting (EH) performance.

\begin{figure}
\centering
  \begin{center}
    \includegraphics[width=0.9\textwidth]{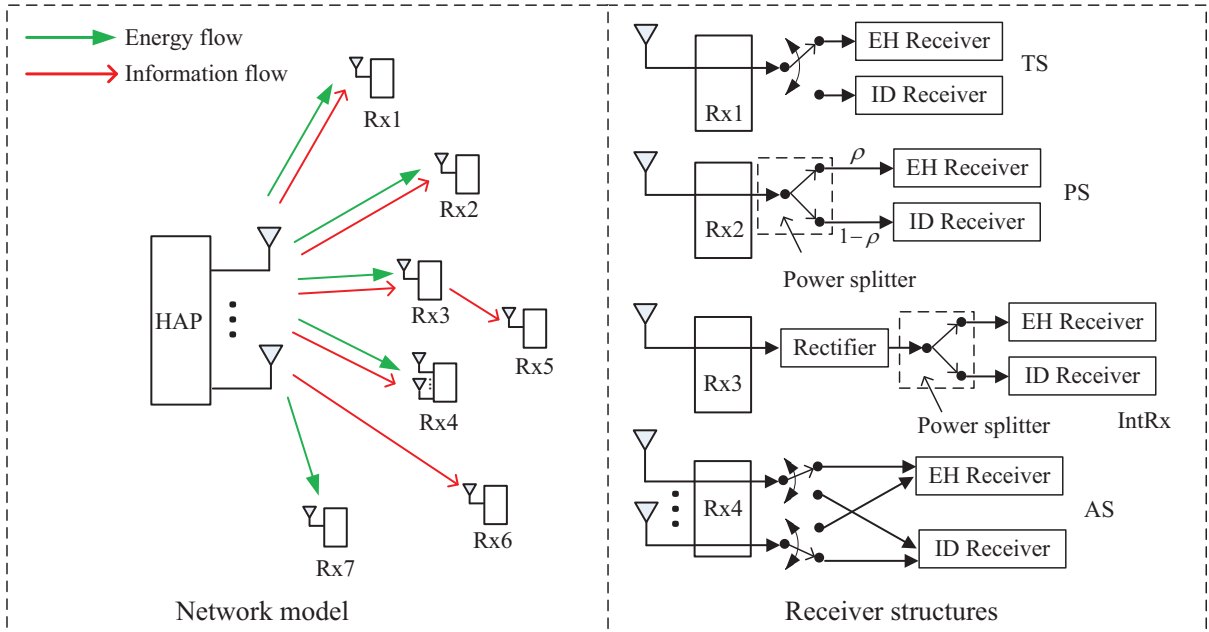}
  \end{center}
  \caption{A SWIPT network model and the receiver structures.}
  \label{74}
\end{figure}

\subsection{Rate-energy tradeoff}
In Fig.~$\ref{74}$, we illustrate a SWIPT network with a multi-antenna hybrid access point (HAP) transmitting energy and information jointly to multiple receivers (Rxs). Some of the receivers receive only information (Rx $6$) or only harvest energy (Rx $7$), while some harvest both energy and receive information simultaneously (Rx $1-4$). It is worth pointing out that a typical ID receiver and EH receiver operate with rather different power sensitivities (e.g., $-10$ dBm for EH receivers versus $-60$ dBm for ID receivers). Therefore, EH receivers are in general closer to the transmitter than ID receivers for effective energy reception.

At the transmitter side, the waveforms generated by the HAP directly determine the performance of information and energy transfer. In the extreme case, the HAP could ignore the energy (information) receivers and optimize the waveforms only to maximize the information (energy) transmission efficiency. However, due to the fundamental difference in the optimal waveforms for information and energy transmissions, such an off-balance design may lead to poor performance of either information or energy transmission. In general, the waveform design needs to follow a \emph{rate-energy tradeoff} to achieve the best possible balance between the two objectives \cite{2013:Zhang}.

Meanwhile, the characterization of rate-energy tradeoff is closely related to the receiver structure and the corresponding signal processing strategies \cite{2013:Zhang,2013:Zhou}. An ideal SWIPT receiver is assumed to be able to decode information and harvest energy from the same signal \cite{2008:Varshney}; however, this could not be realized by practical circuits. Some practical receiver structures are plotted in Fig.~$\ref{74}$, namely, time switching (TS), power splitting (PS), integrated ID/EH receiver (IntRx), and antenna switching (AS) \cite{2013:Zhang,2013:Zhou}, and will be specified in the sequel along with the respective rate-energy tradeoff characterization.

\subsection{Practical receiver structures}
For the simplicity of illustration, we consider each pair of transmitter and receiver separately to discuss the rate-energy tradeoff in a point-to-point channel. In this case, the rate-energy tradeoff is often characterized by the boundary of the \emph{rate-energy region}, defined as the union of all the achievable rate-energy pairs by the receiver. Then, any point on the boundary specifies the maximum achievable data rate given a harvested energy requirement.

\subsubsection{Time switching (TS) receiver} This corresponds to Rx $1$ in Fig.~$4$. A TS receiver consists of co-located ID and EH receivers, where the ID receiver is a conventional information decoder and the EH receiver's structure follows that in Fig.~$\ref{73}$. In this case, the transmitter divides the transmission block into two orthogonal time slots, one for transferring power and the other for transmitting data. At each time slot, the transmitter could optimize its transmit waveforms for either energy or information transmission. Accordingly, the receiver switches its operations periodically between harvesting energy and decoding information between the two time slots. Then, different R-E tradeoffs could be achieved by varying the length of energy transfer slot.

\subsubsection{Power splitting (PS) receiver} This corresponds to Rx $2$ in Fig.~$4$. The EH and ID receiver components of a PS receiver are the same as those of a TS receiver. However, the HAP cannot optimize transmitted signals only for information or energy. Instead, the PS receiver splits the received signal into two streams, where one stream with power ratio $0\leq \rho\leq 1$ is used for EH and the other with power ratio $\left(1- \rho\right)$ is used for ID. Different R-E tradeoffs are achieved by adjusting the value of $\rho$.

\subsubsection{Integrated receiver (IntRx)} This corresponds to Rx $3$ in Fig.~$\ref{74}$. Unlike the TS and PS receivers that split the signal at the RF band, an IntRx combines the RF front-ends of ID and EH receivers and splits the signal after converting it into DC current. Then, the DC current is divided into two streams for battery charging and information decoding, respectively. IntRx uses a (passive) rectifier for RF-to-baseband conversion, which saves the circuit power consumed by the active mixer used in the information decoder of TS/PS receivers. However, the ID receiver of IntRx needs to perform noncoherent detection from the baseband signal (DC current). In this case, conventional phase-amplitude modulation (PAM) must be replaced by \emph{energy modulation}, where information is only encoded in the power of the input signal resulting a reduction of capacity \cite{2013:Zhou}. However, IntRx is superior than the PS/TS receivers when more harvested energy is required, because active frequency down conversion is not performed.

\begin{figure}
\centering
  \begin{center}
    \includegraphics[width=0.6\textwidth]{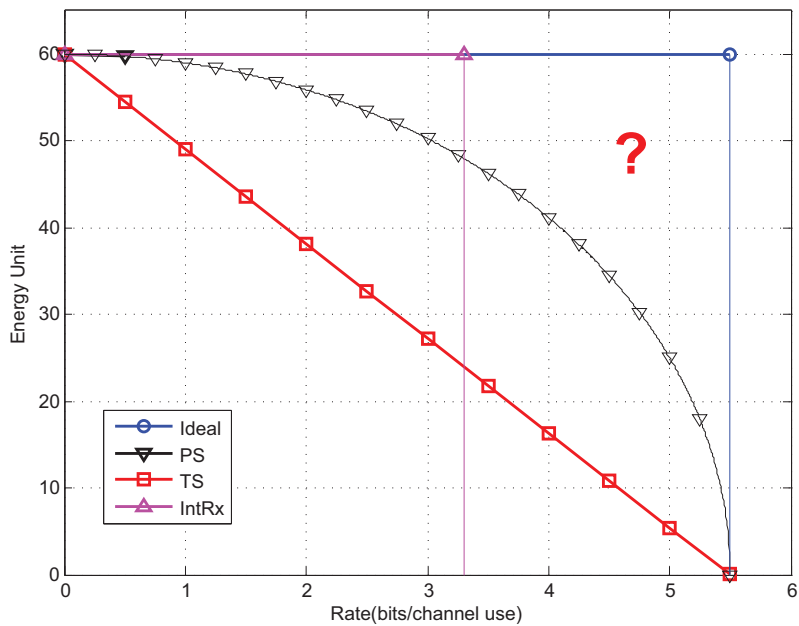}
  \end{center}
  \caption{Comparison of rate-energy tradeoffs of SWIPT receivers.}
  \label{75}
\end{figure}

In Fig.~$\ref{75}$, we give an example to illustrate the key characteristics of the rate-energy regions of the three practical receivers and the ideal receiver in a point-to-point additive white Gaussian noise (AWGN) channel. We could see that, the rate-energy region of the ideal receiver is a box, thus no design tradeoff is involved in this case. A similar rate-energy region is also observed for IntRx. This is because the optimal strategy is to use infinitesimally small amount of DC current for ID and the remaining DC current for EH. The rate-energy region of TS receiver is a straight line connecting the two optimal operating points for EH and ID, respectively. Compared to TS receiver, PS receiver has a strictly larger rate-energy region. So far, the optimal EH-ID receiver is not known. It is unclear if the non-trivial rate-energy region between the ideal and the introduced practical receivers could be achieved or not, which is left for future exploration likely involving different domains, such as physics, circuit theory and information theory.

\subsection{Design challenges and opportunities}
Channel fading and interference are two major challenges to the transceiver design of wireless communication. However, they result in quite different performance degradation for ID and EH in SWIPT. While deep channel fading degrades the performance of both ID and EH, strong interference is only harmful to ID but in fact could be helpful to increase the harvested energy for EH receiver. To optimize the EH and ID performance, the receiver could adapt its strategy to the channel conditions and interference level \cite{2013:Liu1}. For instance, the TS Rx $1$ in Fig.~$4$ should switch to perform ID when the received signal (information and interference) is relatively weak and the signal-to-noise ratio (SNR) is sufficiently high, and EH otherwise. Similarly, the PS Rx $2$ should allocate more received power to ID receiver when the channel is in poor condition, and more power to EH receiver otherwise \cite{2013:Liu2}. Intuitively, this is because the gain achieved by ID receiver is less than the gain achieved by EH receiver when the interference is strong (harmful for ID but helpful for EH), and when the channel is in good condition (logarithmic increase in rate for ID but linear increase in energy for EH). When CSIT is available, the transmitter could also adapt its transmit power to the channel state to achieve the maximum information and energy transfer efficiency, e.g., it does not waste energy to transmit in the case of deep channel fading \cite{2013:Liu1,2013:Liu2}.

The application of MIMO technology could significantly mitigate the effect of channel fading for both energy transmission (energy beamforming) and information transmission (spatial diversity and/or multiplexing) \cite{2013:Zhang}. In a broadcast channel such as in Fig.~$\ref{74}$, a multi-antenna HAP could utilize the spatial degrees of freedom to focus the antenna radiation at specific locations, which not only enhances the harvested energy but also mitigates the interference to unintended information receivers. At the receiver side, the use of multiple antennas enables a low-complexity implementation for PS, named \emph{antenna switching} (see Rx $4$ in Fig.~$\ref{74}$), which uses a subset of antennas for EH ($\rho= 1$) and the rest for ID ($\rho= 0$). While PS requires a power splitter for each antenna, AS reduces the hardware complexity by simply connecting an antenna to either an ID receiver or an EH receiver with an inexpensive switch. The rate-energy region of AS approaches that of a PS receiver when the number of receive antennas is large enough \cite{2013:Liu2}.

SWIPT could also be extended to other useful application scenarios. In Fig.~$\ref{74}$, for instance, the HAP could broadcast energy to the nearby receivers (Rx $7$) and transmit information to the faraway receivers (Rx $6$) simultaneously, while meeting the different sensitivities of EH and ID receivers. Besides, information secrecy could be achieved between the HAP and information receivers using physical layer secrecy coding techniques \cite{2014:Liu}. In addition, a relay node Rx $3$ in Fig.~$\ref{74}$ could harvest energy and receive the message dedicated to Rx $5$ that is located further away from the HAP, and then forward the message to Rx $5$ in another time slot to extend the coverage of the HAP.

\section{Wireless Powered Communication Network}
In SWIPT, the wireless devices use the harvested energy to decode the information sent to them. Here, we consider another scenario that the wireless devices use the harvested energy to transmit information. This communication architecture is referred to as wireless powered communication network (WPCN) \cite{2014:Ju1}. In this section, we introduce the basic operations of a WPCN, the key design challenges and solutions, and interesting extensions to many practical network models.

\begin{figure}
\centering
  \begin{center}
    \includegraphics[width=0.6\textwidth]{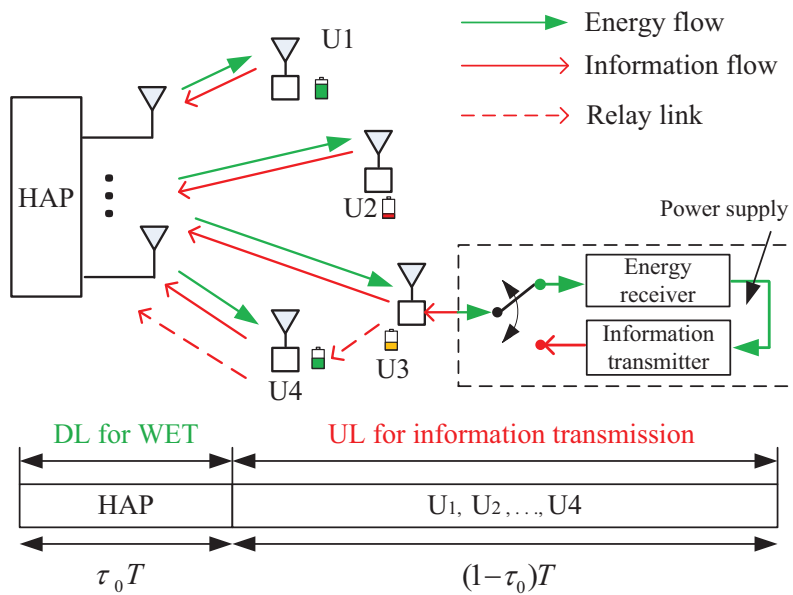}
  \end{center}
  \caption{A WPCN model and the harvest-then-transmit protocol.}
  \label{76}
\end{figure}

\subsection{Harvest-then-transmit protocol}
We consider a single-cell WPCN in Fig.~$\ref{76}$, where a HAP broadcasts energy to the multiple wireless devices in the DL, while the wireless devices communicate to the HAP in the UL using the energy harvested. Due to the half-duplex hardware constraint of HAP, the network operates under a two-phase \emph{harvest-then-transmit} protocol within a transmission block of duration $T$. Specifically, the wireless devices harvest energy from DL WET in the first phase for $\tau_0 T$ ($0<\tau_0<1$) amount of time, and then transmit data in the second phase for the rest of the transmission block. This could be easily achieved by the time-switching circuit model shown for U$3$ in Fig.~$\ref{76}$. Intuitively, with a larger $\tau_0$, the UL data rate could be improved as the devices could harvest more energy in the first phase to transmit data. However, a larger $\tau_0$ also decreases the data rate as it leaves a shorter data transmission time. In general, the optimal value $\tau_0$ that results in the highest UL throughput is related to the users' wireless channel conditions. If the users are all close to the HAP, the optimal $\tau_0$ is small, as each user could still harvest sufficient amount of wireless energy within the short duration of DL WET. Otherwise, a larger $\tau_0$ is required for the far users to harvest sufficient energy before commencing reliable data transmission. In fact, it is shown in \cite{2014:Ju1} that the optimal value of $\tau_0$ that maximizes the sum-rate decreases as the sum of channel power gains of all users increases.

\subsection{Doubly-near-far problem}
Besides setting the optimal duration for WET in the first phase, another important issue in the harvest-then-transmit protocol is to design efficient multiple access scheme in the second phase for coordinating UL information transmission of users. In conventional wireless system, users that are far away from the base station in general achieve lower data rates than those in the vicinity. This fairness issue is even more critical and challenging in a WPCN. Due to significant signal attenuation, a user far away from the HAP (U$2$) harvests much lower wireless energy in the DL but consumes more to transmit data in the UL than a user near the HAP (U$1$). This coupled effect is referred to as the \emph{doubly-near-far problem} \cite{2014:Ju1}, which could result in very low throughput for far users, e.g., $100$ times less data rate than a nearby user, if the multiple access scheme is not properly designed.

When TDMA is used in the second phase, the HAP could allocate a longer data transmission time to the far users in the second phase to tackle the doubly-near-far problem. On the other hand, when the HAP has multiple antennas, \emph{spatial division multiple access} (SDMA) could be applied in the UL. In this case, all the users transmit simultaneously to the HAP during the second phase and the HAP jointly decodes the user messages using multi-user detection (MUD) techniques. SDMA in general achieves higher spectrum efficiency than the TDMA-based method. Besides, the doubly-near-far problem could be mitigated by user transmit power control in the UL and energy beamforming design in the DL. Specifically, the HAP uses EB to steer stronger energy beams towards the far users, and allows them to transmit with a higher power than the near users to balance the throughput performance among all the users \cite{2014:Liu2}.

Another effective method to tackle the doubly-near-far-problem is through user cooperation. In Fig.~$\ref{76}$, after harvesting energy in the first phase, a nearby user U$4$ uses part of its resource (transmit energy and time) in the second phase to forward a far-away user's (U$3$) messages to the HAP \cite{2014:Ju3}. The more resource U$4$ consumes for helping U$3$, the more throughput improvement could be achieved for the far-away user. Due to the transmit time and energy constraints, the relay node U$4$ needs to carefully allocate the resource on relaying the other's message and transmitting its own message. Interestingly, it is shown in \cite{2014:Ju3} that both users could benefit from the cooperation. For the far-away user, the reason is obvious as the cooperation essentially increases the time and energy used for its message transmission. For the nearby user, its data rate loss due to cooperation could be made up by an overall longer data transmission time, because the gain from user cooperation allows the HAP to allocate more time for data transmission, instead of WET.

\subsection{Extensions}
The efficient operation of a WPCN is highly dependent on the accurate knowledge of channel state information (CSI) at the HAP, where both information decoding and resource allocation require accurate CSI estimate. Similar to a conventional wireless network, the throughput performance of WPCN would benefit from a longer channel estimation period with more accurate CSI estimate, but also suffer from a shorter energy/information transmission time. However, a WPCN-specific design tradeoff arises due to the energy constraints at the wireless devices. This is because the wireless devices consume energy for channel estimation on decoding the pilot signals sent by the HAP, transmitting the CSI feedback, or sending pilot signals to the HAP in some channel estimation schemes that exploit the UL/DL channel reciprocity \cite{2014:Zeng}. Evidently, more energy consumption on channel estimation would compromise the transmission rate (or reliability) because less energy is left for communication. However, this results in more accurate CSI estimate and hence more precise beamforming that both improves its transmission rate (or reliability) and increases the harvested energy in the following transmission blocks. The energy of wireless devices for channel estimation and information transmission should thus be carefully allocated to achieve optimal performance.

Another performance enhancing technique to WPCN is massive MIMO, which employs a large number of antennas (tens to more than a hundred) at the HAP to exploit the high antenna array gain. On one hand, the large degree of freedom provided by massive MIMO enables spatial multiplexing to serve more mobile users for UL information transmission at the same time. On the other hand, the large antenna array also improves the EB performance in the DL, by the generation of very sharp beams to enhance the received signal power (e.g., more than $20$ dB power gain). The application of massive MIMO to WPCN could result in multiple folds of throughput improvement and also much longer operating range. Interestingly, the high number of antennas at HAP does not necessarily translate to high processing complexity and cost. The property of asymptotic orthogonality of large user-to-HAP channels leads to largely simplified beamforming design, multiple access control and power control solutions \cite{2014:Yang1}.

More generally, there could be a dedicated wireless energy network consisting of multiple power nodes that broadcast energy by means of WET. Moreover, multiple information receivers may decode the data transmitted by the wireless devices in the UL. In such a multi-cell WPCN (i.e., multiple energy transmitters and information receivers), it is complicated and often intractable to derive the network capacity using the resource allocation methods for single-cell capacity analysis. Instead, the method of \emph{stochastic geometry} is widely used to study the scaling laws of network capacity as a function of system parameters. In the context of a cellular WPCN, \cite{2014:Huang1} proposes to install wireless energy transferring nodes, named power beacons (PBs), to provide energy for the mobile users to transmit data to some base stations (BSs). Based on a stochastic geometry model, it derives the functional relationships between the densities of BSs and PBs as well as their transmission power to achieve a prescribed communication outage probability. The application of WPCN is also exploited in cognitive radio network in \cite{2014:Lee1}, where a secondary transmitter (e.g., sensor) could harvest energy from a primary transmitter (e.g., mobile phone) if they are close enough, and transmit to its intended secondary receiver if it is sufficiently far from any primary transmitter, to avoid potential interference to the primary network.

Other than the extensions discussed above, WPCN could also be applied to many wireless systems with energy constrained wireless devices. For instance, multi-hop communications with energy harvesting relays, systems with densely deployed HAPs using millimeter-wave technologies, and distributed antenna systems with coordinating energy/information beamforming, etc.

\section{Future Research Directions}
WPC contains rich research problems of important applications yet to be studied. In this section. we highlight several interesting research topics we deem particularly worth investigating.

\subsection{Energy and information transfer coexistence}
Due to the critical power constraints of wireless devices, the future wireless system is expected to be a mixture of wireless energy and communication networks. Under spectrum scarcity, it is likely that the two networks operate on overlapped spectrum. This raises the problem for wireless energy and communication networks to coexist. Unlike the two-way interference in conventional multi-cell communication systems, the interference is one-way from energy network to communication network. Besides, the highly different sensitivity of information and energy receivers indicates that the interference due to WET is in general much stronger than information signals. There are many promising solutions to mitigate the interference, such as information/energy transfer scheduling, energy beamforming design, and opportunistic WET based on spectrum sensing, etc. In particular, cognitive radio technology could be used to carry out effective spectrum sensing to minimize the interference from WET to communication network.

\subsection{Cross-layer design}
So far we mainly focus on the physical (PHY) layer techniques to optimize the performance of WPC. In a practical system, medium access control (MAC) plays the key role in determining the fairness and efficiency of the system. An efficient wireless system design often takes a cross-layer approach, especially for the closely related PHY and MAC layers. In the context of WPC, an example of cross PHY-MAC design is for the HAP to steer the energy beam towards a user with relatively strong wireless channel and many data packets backlogged in the queue, rather than considering the physical channel condition alone. Besides, an efficient energy scheduling should also consider the residual battery life, the wake up/sleep schedule, and the expected energy consumptions of all wireless devices.

\subsection{Hardware implementation}
The current studies on WPC are mainly theoretical in nature. The achievable throughput performance using off-the-shelf energy harvesting and communication modules and under practical wireless environment is not known. Hardware prototyping is urgently needed to evaluate the feasibility of WPC, and to test the applications of various technologies in joint energy and information transmissions, such as massive MIMO, millimeter wave, and distributed antenna system. An extensive testbed could also help identify the most suitable technology, and the proper application scenarios for WPC.

\subsection{Health and safety}
With the potential use of massive MIMO and advanced beamforming technologies, the intensity of microwave at a particular area could be strong enough to harm the human health and cause safety issues. In practice, the radiation power of any wireless device must satisfy the equivalent isotropically radiated power (EIRP) requirement on its operating frequency band, e.g., the FCC (Federal Communications Commission) permits a maximum $36$ dBm EIRP on $2.4$ GHz band. One promising method to solve the safety problem is to use distributed antenna system, such that for each antenna the radiation is omnidirectional and relatively weak (thus satisfying the given EIRP constraint), while the combined effect is constructive only at the destined location but destructive at almost everywhere else. This will reduce the risk of ``radiation burn" due to human blockage from a random direction. Besides, the distributed antenna system could be combined with advanced sensing technology to detect the presence of human in real-time, and cease energy transmission if it deems the transmission to be harmful.

\section{Conclusions}
In this article, we have provided an overview of state-of-the-art RF-enabled WET technologies and their applications to wireless communications. Promisingly, wireless powered communications could significantly improve upon its battery-powered counterpart, and be practically achieved using simple and inexpensive transceiver structures. The opportunities and challenges in the design of wireless powered communications were demonstrated by studying two new paradigms: SWIPT and WPCN. We hope that the design of wireless powered communications will spur research innovations in wireless technologies, as the future wireless systems is expected to be a mixture of wireless information and energy transfer, with RF-enabled WET, SWIPT and WPCN being important building blocks.

\newpage 
\begin{IEEEbiographynophoto}{Suzhi Bi}
(bsz@nus.edu.sg) received the B.Eng. degree in communications engineering from Zhejiang University, Hangzhou, China, in 2009. He received the Ph.D degree from The Chinese University of Hong Kong, Hong Kong, in 2013. He is currently a research fellow in the Department of Electrical and Computer Engineering at National University of Singapore, Singapore. His current research interests include wireless information and power transfer, medium access control in wireless networks, and smart power grid communications.
\end{IEEEbiographynophoto}

\begin{IEEEbiographynophoto}{Chin Keong Ho}
(hock@i2r.a-star.edu.sg) received the B.Eng. and M. Eng degrees from National University of Singapore. He received the Ph.D. degree from Eindhoven University of Technology, The Netherlands, where he concurrently conducted research work in Philips Research. He is Lab Head of Energy-Aware Communications Lab in Institute for Infocomm Research, A$\*$STAR. His research interest includes green wireless communications with focus on energy-efficient solutions and with energy harvesting constraints, and implementation aspects of multi-carrier and multi-antenna communications.
\end{IEEEbiographynophoto}

\begin{IEEEbiographynophoto}{Rui Zhang}
(elezhang@nus.edu.sg) received the B.Eng. and M.Eng. degrees from National University of Singapore and the Ph.D. degree from Stanford University, all in electrical engineering. He is now an Assistant Professor with the Department of Electrical and Computer Engineering at National University of Singapore. His current research interests include multiuser MIMO, cognitive radio, energy efficient and energy harvesting wireless communication, and wireless information and power transfer.
\end{IEEEbiographynophoto}

\end{document}